\documentstyle[aps,eqsecnum,preprint]{revtex}
\begin{document}
\preprint{MSUCL-1004}

\title{
The Dynamical Structure of the $\Delta$-Resonance\\
and\\
its Effect on Two- and Three-Nucleon Systems
}

\author{G. Kortemeyer}
\address{
Institut f\"ur Theoretische Physik, Universit\"at Hannover, 
D-30167 Hannover, Germany\\
and\\
NSCL/Cyclotron Laboratory, Michigan State University, East Lansing, MI
48824-1321, U.S.A\cite{GKNow}}
\author{M. T. Pe\~na}
\address{Centro de Fisica Nuclear da Universidade de Lisboa, P-1699 Lisboa
Codex, Portugal\\
and\\
Instituto Superior Technico, Lisboa, P-1096 Lisboa Codex, Portugal}
\author{P. U. Sauer}
\address{
Institut f\"ur Theoretische Physik, Universit\"at Hannover, 
D-30167 Hannover, Germany}
\author{A. Stadler}
\address{College of William and Mary, Williamsburg, VA 23185,
U.S.A.\cite{ASNow}}
\maketitle

\begin{abstract}
The pion-nucleon interaction in the $P_{33}$ partial wave is assumed to proceed
simultaneously through the excitation of the $\Delta$-isobar and through a
phenomenologically introduced non-resonant background potential. The
introduction of the background potential allows a more realistic
parameterization of the pion-nucleon-$\Delta$ vertex compared with the
previously used one without background. It also modifies the propagation of the
$\Delta$-isobar in the nuclear medium and gives rise to novel effective
nucleon-$\Delta$ interactions. Their consequences on predictions for
observables in the two-nucleon system at intermediate energies and in the
three-nucleon bound state are studied.
\end{abstract}

\narrowtext
\section{Introduction}
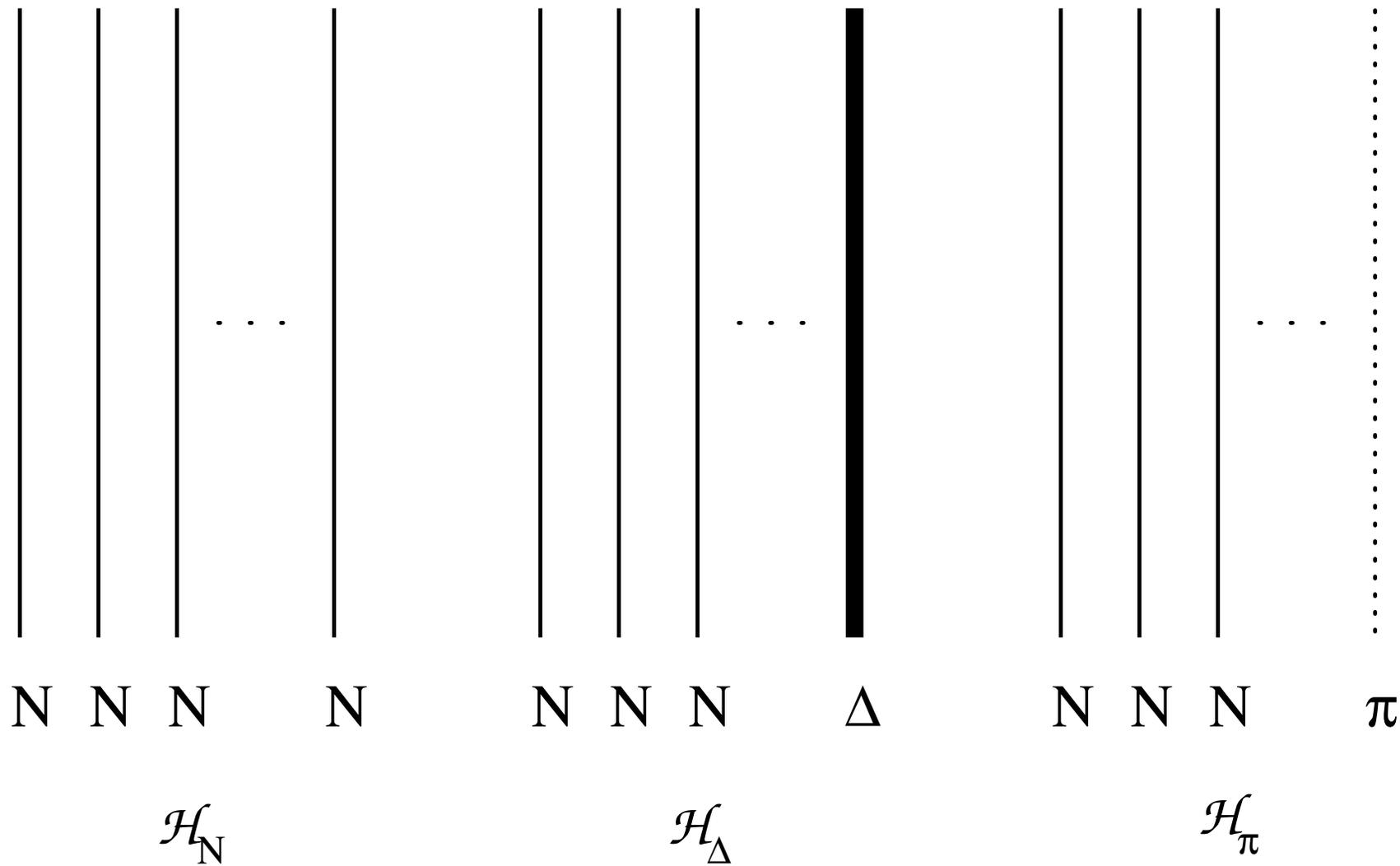
\begin{figure}
\caption[]
{Hilbert space for the description of nuclear phenomena at intermediate
energies. It consists of three sectors: The sector ${\cal H}_N$ contains purely
nucleonic states; in ${\cal H}_\Delta$ one nucleon is turned into a 
$\Delta$-isobar; in ${\cal H}_\pi$ one pion is added. Nucleons will be denoted
by narrow solid lines, $\Delta$-isobars by thick solid lines
and pions by dotted lines.\label{fig1}}
\end{figure}

Internal nucleonic degrees of freedom can get excited when nucleons 
interact. The lowest state of nucleonic excitation is the $\Delta$;
it decays into pion-nucleon ($\pi N$) states. Thus, in nuclear phenomena at
intermediate energies $\Delta$-isobar and pion degrees of freedom become
active. A hamiltonian describing hadronic and e.m. processes at intermediate
energies has to take those degrees of freedom explicitly into account. The
Hilbert space to be considered is shown in Fig.~\ref{fig1}.
Besides the nucleonic sector ${\cal H}_N$, it contains a sector 
${\cal H}_\Delta$ with one nucleon turned into a
$\Delta$-isobar and a sector ${\cal H}_\pi$ with one pion added to the 
nucleons;
their projectors are denoted by $P_N$, $P_\Delta$ and $Q$, respectively.
The hamiltonian $H=H_0+H_1$, to be used to describe the hadronic properties of
the two-baryon system with the inclusion of pion production and pion
absorption, is diagrammatically defined in Fig.~\ref{fig2}; $H_0$ denotes its
kinetic part, $H_1$ its interaction.

The $\Delta$-{\it isobar}, which is introduced in Fig.~\ref{fig1} in the 
Hilbert sector ${\cal H}_\Delta$, is a fictitious baryon of positive parity, 
spin $\frac32$ and isospin $\frac32$. A fixed real mass, which is a parameter 
of the model, and a vanishing width are assigned to it. The $\Delta$-isobar is
unobservable, cross
sections leading to it are identically zero. In contrast, the physical
{\it resonance} at 1232 MeV in the $P_{33}$ partial wave of 
pion-nucleon scattering is composed of 
$\Delta$-isobar and $\pi N$ states in the model. 
The hamiltonian models the resonance, which has the 
physical properties of an 
effective energy-dependent mass and a non-vanishing energy-dependent 
width\cite{1}, through the
$\pi N\Delta$ vertex $QH_1P_\Delta$ of Fig.~\ref{fig2}(e) and 
the $\pi N$ potential
$QH_1Q$ of Fig.~\ref{fig2}(f).

The modelling of the $P_{33}$ resonance by the hamiltonian is non-unique.
E.g., $\pi N$ scattering up to a mass of 1500 MeV can fully be accounted for
under the assumption of a vanishing $\pi N$ background potential, i.e., with
$QH_1Q=0$. In this case, a mass parameter of $m_\Delta^0=1311$ MeV$/c^2$ is
assigned to the $\Delta$-isobar; the regularizing cutoff mass for the $\pi
N\Delta$ vertex is very small with $\Lambda=288$ MeV; as a consequence,
self-energy corrections of the $\Delta$-isobar in the nuclear medium turn out
to be quite moderate. The model for the $P_{33}$ resonance without non-resonant
background has been used by the authors in the past \cite{1,3,4,2}. This paper
introduces an alternative parameterization and explores its consequences for the
description of two- and three-nucleon systems: The $\pi N$ background potential
$QH_1Q$ is assumed to be non-vanishing; processes arising from two-pion
channels and from meson exchange between pion and nucleon contribute to the
background; for simplicity, however, we choose to parametrize the background in
a separable form. Furthermore, we require the regularizing cutoff mass 
$\Lambda$ for the $\pi N\Delta$ vertex to be of the order of 1 GeV, a 
magnitude familiar from
realistic one-boson exchange two-nucleon potentials; but within that order of
magnitude the cutoff mass $\Lambda$ remains a fit
parameter. The fit to $P_{33}$ $\pi
N$ phase shifts determines the parameters in the one-baryon part of the
hamiltonian $H$. E.g., the mass parameter of the $\Delta$-isobar
becomes with $m_\Delta^0=1801$ MeV$/c^2$ quite different from the value for the
resonance position of the physical $P_{33}$ resonance. Thus, self-energy
corrections of the $\Delta$-isobar in the nuclear medium get dramatically large
as will be demonstrated later on.

Sect.~\ref{sec2} describes the two different models for the $P_{33}$ $\pi N$
resonance. Different parameterizations for the hamiltonian $H$ of
Fig.~\ref{fig2} result. Consequences arising from the different
parameterizations of the hamiltonian on predictions for properties of
the two-nucleon system above pion threshold are explored in Sect.~\ref{sec3};
consequences for the three-nucleon bound state are explored in Section
\ref{sec4}. Sect.~\ref{sec5} sums up the conclusions.

\section{Models for the $P_{33}$ Pion-Nucleon Resonance}\label{sec2}
This section describes $P_{33}$ $\pi N$ scattering in the framework of the
hamiltonian defined in Fig.~\ref{fig2}. It assumes that the $\pi N$
background potential of Fig.~\ref{fig2}(f) may not be zero.
The considered $\pi N$ hamiltonian has the following parts, i.e., the 
kinetic energy $H_0$, the $\pi N\Delta$ vertex $QH_1P_\Delta$ and the 
$\pi N$ potential
$QH_1Q$. The one-baryon nature of the operators is made explicit by
the notation
\begin{eqnarray}
H_0&=&\sum_i\left[P_Nh_0(i)P_N+P_\Delta
h_0(i)P_\Delta+Qh_0(i)Q\right]\nonumber\\
&&+Qh_0(\pi)Q\ ,\label{2.1}
\end{eqnarray}
\begin{mathletters}
\begin{eqnarray}
QH_1P_\Delta&=&\sum_iQh_1(i)P_\Delta,\\
 P_\Delta H_1Q&=&[QH_1P_\Delta]^{\dag}\ ,
\end{eqnarray}
\end{mathletters}
\begin{equation}
QH_1Q=\sum_iQh_1(i)Q\quad\label{2.3}
\end{equation}
as in Ref.~\cite{2}. The index $i$ denotes the baryon that the respective
operator acts on, i.e., the baryon $i$ in the kinetic energy operator, the 
$\Delta$-isobar
in $QH_1P_\Delta$, and the nucleon involved in the $\pi N$ background
interaction. 

Eqs.~(\ref{2.1})-(\ref{2.3}) remind us that the operators corresponding to
Fig.~\ref{fig2} are defined in the Hilbert space of two baryons; in the
reduction to the one-baryon process of $\pi N$ scattering the label $i$ will be
omitted. The form (\ref{2.3}) of $QH_1Q$ is not general, in contrast to the
$\pi N$ background potential, we still assume the $NN$ potential of
Fig.~\ref{fig2}(g) in the Hilbert sector ${{\cal H}}_\pi$ to be vanishing.
This assumption has no consequences in $\pi N$ scattering. 
However, in two- and three-baryon systems, it is a physics approximation,
that Ref.~\cite{nn4} finds to be minor.

The hamiltonian yields the following $\pi N$ transition matrix $Qt(z)Q$ in the
$P_{33}$ partial wave,
\widetext
\begin{mathletters}
\begin{eqnarray}
\nonumber Qt(z)Q&=&Qt_{BG}(z)Q+\left[1+Qt_{BG}(z)Q\frac Q{z-Q(h_0+h_0(\pi))Q}
\right]\\
\nonumber&\times&Qh_1P_\Delta\frac{P_\Delta}{z-P_\Delta h_0P_\Delta-P_\Delta
h_1Q\displaystyle\frac Q{z-Q(h_0+h_0(\pi)+h_1)Q}Qh_1P_\Delta}P_\Delta h_1Q\\
&\times&\phantom{Qt_{BG}(z)Q+}
\left[1+Qt_{BG}(z)Q\frac Q{z-Q(h_0+h_0(\pi))Q}\right],\\
Qt_{BG}(z)Q&=&Qh_1Q\left[1+\frac Q{z-Q(h_0+h_0(\pi))Q}Qt_{BG}(z)Q\right].
\end{eqnarray}
\end{mathletters}
The resulting transition matrix is a complicated and non-linear superposition
of resonant and non-resonant contributions. We identify as its resonant part
\begin{eqnarray}
\nonumber&&Qh_1P_\Delta\frac{P_\Delta}{z-P_\Delta h_0P_\Delta-P_\Delta
h_1Q\displaystyle\frac Q{z-Q(h_0+h_0(\pi)+h_1)Q}Qh_1P_\Delta}P_\Delta h_1Q\\
&&\qquad=
Qh_1P_\Delta\frac1{z-M_\Delta(z,k_\Delta)c^2+\frac i2\Gamma(z,k_\Delta)-
\displaystyle\frac{\hbar^2k_\Delta^2}{2m_\Delta^0}}P_\Delta h_1Q,
\label{2.5}
\end{eqnarray}
whereas $Qt_{BG}(z)Q$ carries the 
information on the $\pi N$ background potential
$Qh_1Q$. Besides the linear background contribution $Qt_{BG}(z)Q$, the
background generates dressing for the $\pi N\Delta$ vertex and modifies the
$\Delta$-isobar propagator. The operator
$\hbar$\boldmath$k$\unboldmath$_\Delta$ denotes the
$\Delta$-isobar momentum.
Fig.~\ref{fig3} shows characteristic contributions to the transition
matrix $Qt(z)Q$.

\widetext
Eq.~(\ref{2.5}) defines the effective mass and the effective width of the
$\Delta$-isobar needed in $\pi N$ scattering, but also in the nuclear medium,
i.e,
\begin{mathletters}
\begin{eqnarray}
\label{2.6a}
M_\Delta(z,k_\Delta)c^2\ =&m_\Delta^0c^2+&\phantom{2}\mbox{ Re}
\left[P_\Delta h_1Q\frac Q{z-Q(h_0+h_0(\pi)+h_1)Q}Qh_1P_\Delta\right]\ ,\\
\label{2.6b}\Gamma_\Delta(z,k_\Delta)c^2\ =&\qquad -&2\mbox{ Im}
\left[P_\Delta h_1Q\frac Q{z-Q(h_0+h_0(\pi)+h_1)Q}Qh_1P_\Delta\right]\ .
\end{eqnarray}
The effective mass and the effective width depend on the energy $z$ available
for $\pi N$ scattering, on the $\Delta$-isobar momentum
$\hbar$\boldmath$k$\unboldmath$_\Delta$ and on the non-resonant background
$Qt_{BG}(z)Q$, as is obvious due to the standard decomposition
\begin{eqnarray}
&&\nonumber P_\Delta h_1Q\frac Q{z-Q(h_0+h_0(\pi)+h_1)Q}Q h_1P_\Delta=\\
&&P_\Delta h_1Q
\left[\frac Q{z-Q(h_0+h_0(\pi))Q}+\frac Q{z-Q(h_0+h_0(\pi))Q}
Qt_{BG}(z)Q\frac Q{z-Q(h_0+h_0(\pi))Q}\right]Q h_1P_\Delta\ .
\end{eqnarray} 
\end{mathletters}
\narrowtext

This paper employs non-relativistic kinematics in $Qh_0Q$ for the nucleon, but
relativistic kinematics in $Qh_0(\pi)Q$ for the pion.
The $\pi N\Delta$ vertex is parameterized as in \cite{3} to be of the monopole
form
\begin{mathletters}
\begin{eqnarray}
Qh_1P_\Delta&=&|f\rangle\\
\label{2.7b}
\langle k|f\rangle&=&\frac{f^\ast}{m_\pi c}\sqrt{\frac{4\pi}3\frac1{
(2\pi\hbar)^3}}
\frac{\hbar^2 k}{\sqrt{2\omega_\pi(k)}}
\left(\frac{\Lambda^2-m_\pi^2c^2}{\Lambda^2+
\hbar^2k^2}
\right)
\end{eqnarray}
with $f^\ast$ as coupling constant and $\Lambda$ as a regularizing
cutoff momentum.
$m_\pi$ denotes the mass of the pion,
$\omega_\pi(k):=c\sqrt{\hbar^2k^2+m_\pi^2c^2}$ the energy of the pion. 
Instead of the
coupling constant $f^\ast$, the combination
\begin{equation}
\frac{f^2}{4\pi}:=\frac{f^{\ast2}}{4\pi}\left(\frac{\Lambda^2-m_\pi^2c^2}
{\Lambda^2+\hbar^2k^{\ast2}}\right)
\end{equation}
\end{mathletters}
of parameters --- 
$\hbar k^{\ast}$ being the relative $\pi N$ momentum at the resonance
position, e.g., 1232 MeV/$c=m_Nc+\hbar^2 k^{\ast2}/(2m_Nc)+
\omega_\pi(k^{\ast})/c$ with $m_N$
as nucleon mass ---
represents the effective coupling of 0.306 between resonance and $\pi N$ 
states
realistically. The $\pi N$ background potential is chosen to be separable,
i.e.,
\begin{mathletters}
\begin{equation}
Qh_1Q:=\sum_{\alpha=1,2}|g_\alpha\rangle\lambda_\alpha\langle g_\alpha|
\label{2.8}
\end{equation}
with
\begin{equation}
\langle k|g_\alpha\rangle:=\frac{\hbar k^{\ast3}c}{\sqrt{k^\ast}}\frac 
k{(k^2+\beta_\alpha^2)^2}
\ ,
\end{equation}
\end{mathletters}
where $\lambda_\alpha$ and $\beta_\alpha$ are additional free parameters. The
hamiltonian is required to account for the experimental $P_{33}$ $\pi N$ phase
shift in the energy region from threshold to 1500 MeV \cite{5}. 
Table \ref{tab1} summarizes the results (KB) of the fitting procedure for the
parameters $m_\Delta^0$, $\lambda_\alpha$ and $\beta_\alpha$. The
parameterization (P) without background potential on which the calculations
of Refs.~\cite{1,3,4,2} are based is given for comparison; it is adapted to
the parameterization (Pa) by an improved fit in this paper; the adaptation only
yields a minute change in $m_\Delta^0$. Thus, without physics consequences the
parameterizations (P) and (Pa) have been used throughout this paper for
reference purposes.
Fig.~\ref{fig4} shows the good agreement between calculated and measured
phase shifts.

Both descriptions of $P_{33}$ $\pi N$ scattering, i.e., the one without and 
with $\pi N$ background potential, account for
phase shifts with comparable quality. Fig.~\ref{fig4} demonstrates that
differences in the fits are graphically only discernable at larger energies; it
also proves that even in the presence of a background potential the
$\Delta$-isobar provides the dominant contribution to the physical resonance.
The non-resonant component indeed only appears as a background, which justifies
our previous approximation $Qh_1Q=0$ in retrospect. In fact, an expansion of
the $\pi N$ transition matrix in terms of the transition matrix $Qt_{BG}(z)Q$ 
of
the non-resonant potential reproduces phase shifts better than 2\% already in
first order in the region of the resonance within its experimental width. This
is
proof of the comparative weakness of the $\pi N$ background potential; in fact,
the replacement of the background transition matrix by the background
potential, i.e., of $Qt_{BG}(z)Q$ by $Qh_1Q$,
is for the phase shifts an excellent approximation, but becomes considerably
poorer outside the resonance region; the replacement can yield some
deviations for
relative momenta of the $\pi N$ system below 180 MeV/$c$, but stays
within 1\% around the resonance position.  

According to Table~\ref{tab1} the increase of the cutoff momentum $\Lambda$,
responsible for the suppression of the $\pi N\Delta$ vertex with increasing
relative momentum according to Eq.~(\ref{2.7b}), leads to an increase of the 
bare mass $m_\Delta^0$ of the $\Delta$-isobar. 
This correlation reveals a balance: On one hand, the
larger cutoff $\Lambda$ enlarges the coupling of $\pi N$ states to the
$\Delta$-isobar; on the other hand, the larger $\Delta$-mass makes the
same transition energetically less favorable. Despite that balance, a large
bare mass for the $\Delta$-isobar is quite worrisome: It yields substantial 
self-energy corrections for
the $\Delta$-isobar propagation, they are displayed in Fig.~\ref{fig5}. The
variation of the effective $\Delta$-mass $M_\Delta(z,k_\Delta)$ and
$\Delta$-width $\Gamma_\Delta(z,k_\Delta)$ with the available energy $z$ gets
important when the $\Delta$-isobar and the interacting $\pi N$ system are 
imbedded in many-nucleon systems.

\section{Effects on the Two-Nucleon System above Pion Threshold}\label{sec3}
In the two-nucleon system above pion threshold the following processes
involving at most one pion are possible, i.e., $NN\to NN$,
$NN\leftrightarrow\pi d$, $NN\to\pi NN$, $\pi d\to\pi d$ and $\pi d\to\pi NN$;
the symbol $d$ stands for deuteron. The processes are unitarily coupled. The
technique for calculating observables is taken from 
Ref.~\cite{4}; it solves a coupled-channel problem. The coupled channels have
two baryons, either two nucleons or one nucleon and one $\Delta$-isobar. The
transcription into a coupled-channel problem is exact: The channel with a pion
is projected out. However, it signals its presence by an energy-dependent
$N\Delta$ interaction $P_\Delta H_{1\mbox{\scriptsize eff}}(z)P_\Delta$.

Since the pion is produced or absorbed through the $\Delta$, only the
nucleon-$\Delta$ channel receives such effective pionic contributions besides
the instantaneous ones $P_\Delta H_1P_\Delta$
of Fig.~\ref{fig2}. They have the form
\widetext
\begin{mathletters}
\begin{eqnarray}
&&P_\Delta H_{1\mbox{\scriptsize eff}}(z)P_\Delta=P_\Delta H_1P_\Delta
+P_\Delta H_1Q\frac Q{z-Q(H_0+H_1)Q}QH_1P_\Delta,\\
\nonumber&& P_\Delta H_1Q\frac Q{z-Q(H_0+H_1)Q}QH_1P_\Delta\\
&&=\sum_{i,j,k}
P_\Delta h_1(i)Q\left[\frac Q{z-QH_0Q}+\frac Q{z-QH_0Q}Qt_{BG}(z-h_0(k))Q
\frac Q{z-QH_0Q}\right]Qh_1(j)P_\Delta\nonumber\\
&&\qquad\qquad\qquad+{{\cal O}}[(Qt_{BG}(z)Q)^2].\label{3.1b}
\end{eqnarray}
\end{mathletters}
\narrowtext
The arising contributions are displayed in Fig.~\ref{fig6}. They are of
one-baryon and two-baryon nature. The ones without the $\pi N$ background
potentials are shown as processes (a) and (c). Process (b), corresponding to
the part $i=j$, $k\neq i$ in the sum (\ref{3.1b}), modifies the one-baryon
contribution. Process (d), corresponding to $i\neq j$ and $k=j$, and process
(e), corresponding to $i=j=k$, modify the effective $N\Delta$
interaction. The processes (b), (d) and (e) of Fig.~\ref{fig6} are computed in
this paper and added to the corresponding ones without background in the 
formalism
in Ref.~\cite{4} when calculating observables of the two-nucleon system; the
separability (\ref{2.8}) of the background potential $Qh_1Q$ simplifies their
computation a great deal technically.

Eq.~(\ref{3.1b}) is an expansion of the effective $N\Delta$ interaction up to
first order in the $\pi N$ background transition matrix $Qt_{BG}(z)Q$. Only
those first order contributions are retained in the computation; a sample
contribution of second order in $Qt_{BG}(z)Q$, not included, is shown in
Fig.~\ref{fig6}(f). In the case of $P_{33}$ $\pi N$ scattering,
Sect.~\ref{sec2} discussed the validity of such an expansion in powers of
$Qt_{BG}(z)Q$ and found the first order highly satisfactory. It is believed,
though it could not be checked, that the validity carries over to the 
description of the two-nucleon system above threshold. 

The distortion of the asymptotic $\pi d$ states by the background
potential is not considered.

Results for sample observables of elastic two-nucleon scattering, of
pion-production in the two-proton reaction $pp\to\pi^+d$, and of pion-deuteron
scattering are shown in Figs.~\ref{fig7} to \ref{fig9}. The parameterization of
the hamiltonian in the two-baryon system is the same as in Ref.~\cite{4}; it
contains the $N\Delta$ potential based on meson exchange.
The dotted lines in all figures represent the results
for the parameterization (P) of the $\pi N$ interaction without $P_{33}$ 
background potential; the results are only slightly changed with
respect to \cite{4} due to an improvement in calculational technique which is 
described in Ref.~\cite{n4}. The solid
lines represent the results for the new parameterization (KB) containing the
background potential and having a larger 
cutoff momentum $\Lambda$ and a larger bare $\Delta$-mass $m_\Delta^0$; the 
background is included in the propagator
of the $\Delta$-resonance according to Fig.~\ref{fig6}(b), and as a
correction in the pion exchange potential according to Fig.~\ref{fig6}(d)
and Fig.~\ref{fig6}(e); among the latter two corrections, the vertex correction
of Fig.~\ref{fig6}(d) is found to be the much more important one.

Observables sensitive towards changes of the interaction in the Hilbert sector 
${\cal H}_\Delta$ are
phase shifts and inelasticities for the ${}^1D_2$ partial wave in elastic
$NN$-scattering, since the nucleonic ${}^1D_2$ wave is coupled to the
${}^5S_2$ $N\Delta$ wave --- pion production and pion-deuteron scattering.
Fig.~\ref{fig7} shows the ${}^1D_2$ phase shifts.
The dashed line is added to isolate the effect of the background, it
shows the results for (K) of Table~\ref{tab1}, i.e., for 
the new parameterization (KB) while omitting the background contribution. A 
comparison between the
dashed and the solid curve can be used to estimate the direct influence of the
background, as in Fig.~\ref{fig4} for the $P_{33}$ phase shifts;
the $P_{33}$ resonance is sharpened by the changed resonance parameters (KB)
compared with (P), but 
gets broadened by the background potential also in the two-nucleon system.
Fig.~\ref{fig8} shows the influence of the background on differential
cross sections for $NN\to\pi d$, Fig.~\ref{fig9} shows the same for
$\pi d\to\pi d$.

For two energies, computations of the ${}^1D_2$ phase shift are also performed
with the background
potential itself instead of its transition matrix, i.e., for the replacement of
$Qt_{BG}(z)Q$ by $Qh_1Q$. The results
of both computations are found to differ by less than 1\%, a result
that is compatible with the small differences between the two corresponding
calculations of the $\pi N$ phase shifts in Sect.~\ref{sec2}.
Nevertheless, the effect of the background on the considered observables is
quite sizeable. As Fig.~\ref{fig7} proves, the effect is an indirect one; the
background potential changes the bare $\Delta$-mass and the $\pi N\Delta$
vertex parameters, and that change has a large impact on the observables of the
two-nucleon system above pion threshold.

For all considered reactions, the introduction of the $\pi N$ background
potential leads by and large to a poorer
agreement with experimental data except for the ${}^1D_2$ phase shifts. We
attribute this sad fact to the dramatically large self-energy corrections which
the $\Delta$-isobar receives according to Fig.~\ref{fig5}.

\section{Effects on the Three-Nucleon Bound State}\label{sec4}
In the first sections of this paper, the $\Delta$-isobar was used as a reaction
mechanism for pion scattering, pion production and pion absorption; that
reaction mechanism depends on the introduced $\pi N$ background potential in
the $P_{33}$ partial wave. In bound nuclear systems, the explicit
$\Delta$-isobar and pion degrees of freedom yield hadronic and electromagnetic
nuclear-structure corrections compared with a purely nucleonic description,
e.g., effective medium-dependent many-nucleon interactions and currents. As
long as the Hilbert sector ${\cal H}_\pi$ with a pion is assumed to be
interaction-free, i.e., $QH_1Q=0$, the effective many-nucleon interactions and
currents remain reducible into one- and two-baryon contributions. Clearly, the
two-baryon processes of Fig.~\ref{fig6} keep that character even when imbedded
in a larger nuclear medium. However, the $\pi N$ background potential also 
yields three-baryon
contributions which are irreducible in the baryonic Hilbert sectors. 
Fig.~\ref{fig10} shows examples for the effective three-baryon interaction 
which arises in 
the Hilbert sector ${\cal H}_\Delta$. Technically, it can be treated in the 
three-nucleon bound state as any irreducible three-baryon force according to 
the technique of Ref.~\cite{n7}. However, such an exact calculation is 
technically very demanding and may even not be necessary. Sect.~\ref{sec2}
concluded that the $\pi N$ background is weak and can reliably be treated in
perturbation theory. This section developes such an approximation scheme. The
calculations will keep only the two-baryon processes of Fig.~\ref{fig6}.

\begin{mathletters}
\widetext
The three-nucleon bound state $|B\rangle$ satisfies the following
coupled-channel Schr\"odinger equation, i.e.,
\begin{equation}
\left[(P_N+P_\Delta)H(P_N+P_\Delta)+P_\Delta H_1Q\frac Q{E_T-QHQ}QH_1P_\Delta
\right](P_N+P_\Delta)|B\rangle=E_T(P_N+P_\Delta)|B\rangle\ ,\label{4.1a}
\end{equation}
\begin{equation}
Q|B\rangle=\frac Q{E_T-QHQ}QH_1P_\Delta\ P_\Delta|B\rangle\ ,\label{4.1b}
\end{equation}
with the normalization condition
\begin{equation}
\langle B|(P_N+P_\Delta+Q)|B\rangle=1\ .\label{4.1c}
\end{equation}
\end{mathletters}
The exact set of equations (\ref{4.1a})-(\ref{4.1c}) is compared with
the approximate one, in which the $\pi N$ background potential is neglected, 
i.e., $QH_1Q=0$. In zeroth order of
the $\pi N$ background the approximate trinucleon binding energy and wave
function are $E_T^{[0]}$ and $|B^{[0]}\rangle$, respectively. We use the
following steps in order to relate the exact and the approximate eigenvalues
\begin{mathletters}
\begin{equation}
E_T=\frac{\langle B|(P_N+P_\Delta)H(P_N+P_\Delta)+P_\Delta H_1
Q\displaystyle\frac Q{E_T-QHQ}QH_1P_\Delta|B\rangle}
{\langle B|P_N+P_\Delta|B\rangle}\label{4.2a}
\end{equation}
\begin{equation}
E_T\approx\frac{\langle B^{[0]}|(P_N+P_\Delta)H(P_N+P_\Delta)+P_\Delta H_1
Q\displaystyle\frac Q{E_T^{[0]}-QHQ}QH_1P_\Delta|B^{[0]}\rangle}
{\langle B^{[0]}|P_N+P_\Delta|B^{[0]}\rangle}\label{4.2b}
\end{equation}
\begin{equation}
E_T\approx E_T^{[0]}+\frac{\langle B^{[0]}|P_\Delta H_1Q\left[\displaystyle
\frac Q{E_T^{[0]}-QHQ}-\displaystyle\frac Q{E_T^{[0]}-QH_0Q}\right]QH_1P_\Delta
|B^{[0]}\rangle}{\langle B^{[0]}|P_N+P_\Delta|B^{[0]}\rangle}\label{4.2c}
\end{equation}
\end{mathletters}
The background potential $QH_1Q$ is assumed to change the baryonic wave
function components, i.e.,
\begin{mathletters}
\begin{eqnarray}
P_N|B\rangle&\approx&P_N|B^{[0]}\rangle\ ,\label{4.3a}\\
P_\Delta|B\rangle&\approx&P_\Delta|B^{[0]}\rangle\ ,\label{4.3b}
\end{eqnarray}
and the available energy in the effective interaction, i.e.,
\begin{equation}
P_\Delta H_1Q\frac Q{E_T-QHQ}QH_1P_\Delta\approx P_\Delta H_1Q\frac Q{E_T^{[0]}
-QHQ}QH_1P_\Delta\ ,\label{4.3c}
\end{equation}
by very little. The assumptions (\ref{4.3a})-(\ref{4.3c}) yield the step from
Eq.~(\ref{4.2a}) to Eq.~(\ref{4.2b}), and only thereby to Eq.~(\ref{4.2c}).

Since the difference propagator $[Q/(E_T^{[0]}-QHQ)-Q/(E_T^{[0]}-QH_0Q)]$ can
be expanded in powers of the background transition matrix $Qt_{BG}(z)Q$, the
perturbation scheme (\ref{4.2c}) for the binding energy $E_T$ is ordered
according to powers of $Qt_{BG}(z)Q$. 
The perturbation scheme (\ref{4.2c}) does not
follow from the Ritz variational principle which works with an expansion in
powers of the potential $Qh_1Q$. We note, however, that for relevant available
energies $z$ $Qt_{BG}(z)Q\approx Qh_1Q$ as verified in Sect.~\ref{sec2} for
the $\pi N$ $P_{33}$ phase shifts and in Sect.~\ref{sec3} for the 
$NN$ ${}^1D_2$
phase shifts and inelasticities, and that 
$\langle B^{[0]}|(P_N+P_\Delta)|B^{[0]}
\rangle\approx1$; thus, in this approximation the perturbation scheme 
(\ref{4.2c}) becomes 
\end{mathletters}
\begin{equation}
E_T=E_T^{[0]}+\langle B^{[0]}|P_\Delta H_1Q\frac Q{E_T^{[0]}-QH_0Q}QH_1Q
\frac Q{E_T^{[0]}-QH_0Q}QH_1P_\Delta|B^{[0]}\rangle
\end{equation}
and is therefore almost variational.
\narrowtext

This section compares trinucleon results obtained for the two different
parameterizations (KB) and (P) of the $P_{33}$ $\pi N$ resonance with and
without $\pi N$ background potential according to Sect.~\ref{sec2} and
Table~\ref{tab1}. An exact Faddeev calculation is done for (KB) without 
background; we identify its results with $|B^{[0]}\rangle$ and $E_T^{[0]}$ of
Eq.~(\ref{4.2c}). First-order perturbation theory according to 
Eq.~(\ref{4.2c}) is then used to obtain an improved result for the triton 
binding energy. The
perturbation calculations include the one- and two-baryon contributions of 
Figs.~\ref{fig6}(b),
\ref{fig6}(d) and \ref{fig6}(e), all contributions being of first order in the
$\pi N$ background transition matrix $Qt_{BG}(z)Q$; the three-baryon process of
Fig.~\ref{fig10}(a) could not be included. However, the validity of the
perturbation theory, first order in $Qt_{BG}(z)Q$, could be checked by
comparing the perturbative results for the process of Fig.~\ref{fig6}(d) with
an exact Fadeev calculation; we claim agreement between the exact and
perturbative results on the level of numerical
accuracy. Among the three processes the one of Fig.~\ref{fig6}(e) accounts for
less than
1 eV, which is an order of magnitude smaller than the contributions of the other
two, being of the order of a few keV.
  
The obtained results are collected in Table \ref{tab2}. It lists the triton
binding energy $E_T$ and the wave function probabilities $P_{{\cal L}}$,
$P_\Delta$ and $P_\pi$ for the nucleonic components of total angular momentum
${{\cal L}}$ and of particular orbital symmetry, for the components with a
$\Delta$-isobar, and for the components with a pion. The rows 2 and 3 give the
changes in binding energy due to the considered non-nucleonic degrees of
freedom; $\Delta E_2$ is the change due to effective two-nucleon contributions,
$\Delta E_3$ is the change due to effective three-nucleon contributions, as
defined in Ref.~\cite{11}.

The parameterization (KB) of the $\pi N$ interaction with background leads to 
a tiny decrease of the binding 
energy compared with the traditional calculation in Ref.~\cite{11} 
based on the parameterization (P) without background. The decrease 
corresponds to a decrease of both non-nucleonic effects $\Delta E_2$ and 
$\Delta E_3$. Their reduction is
plausible, since for (KB) compared with (P) the energy difference
$(m_\Delta^0-m_N)c^2$ is more than doubled. Thus, the excitation of the
$\Delta$-isobar gets energetically unfavorable. This fact is borne out by
the substantial reduction of the trinucleon $\Delta$-probability $P_\Delta$ 
from 1.71\% to 1.16\%. In contrast, in the parameterization (KB) the decay of 
the
$\Delta$-isobar into $\pi N$-states is less inhibited; this is the reason why
the probability $P_\pi$ of pionic components is increased.
In fact, from the ratio of $P_\Delta$ and $P_\pi$
one can conclude that in the old parameterization (P) the 
$\Delta$-resonance in the  trinucleon system has
about 3\% pionic components, however, in the new parameterization (KB) more 
than 14\%. The
influence of the background on these values is very small.
\section{Conclusions}\label{sec5}
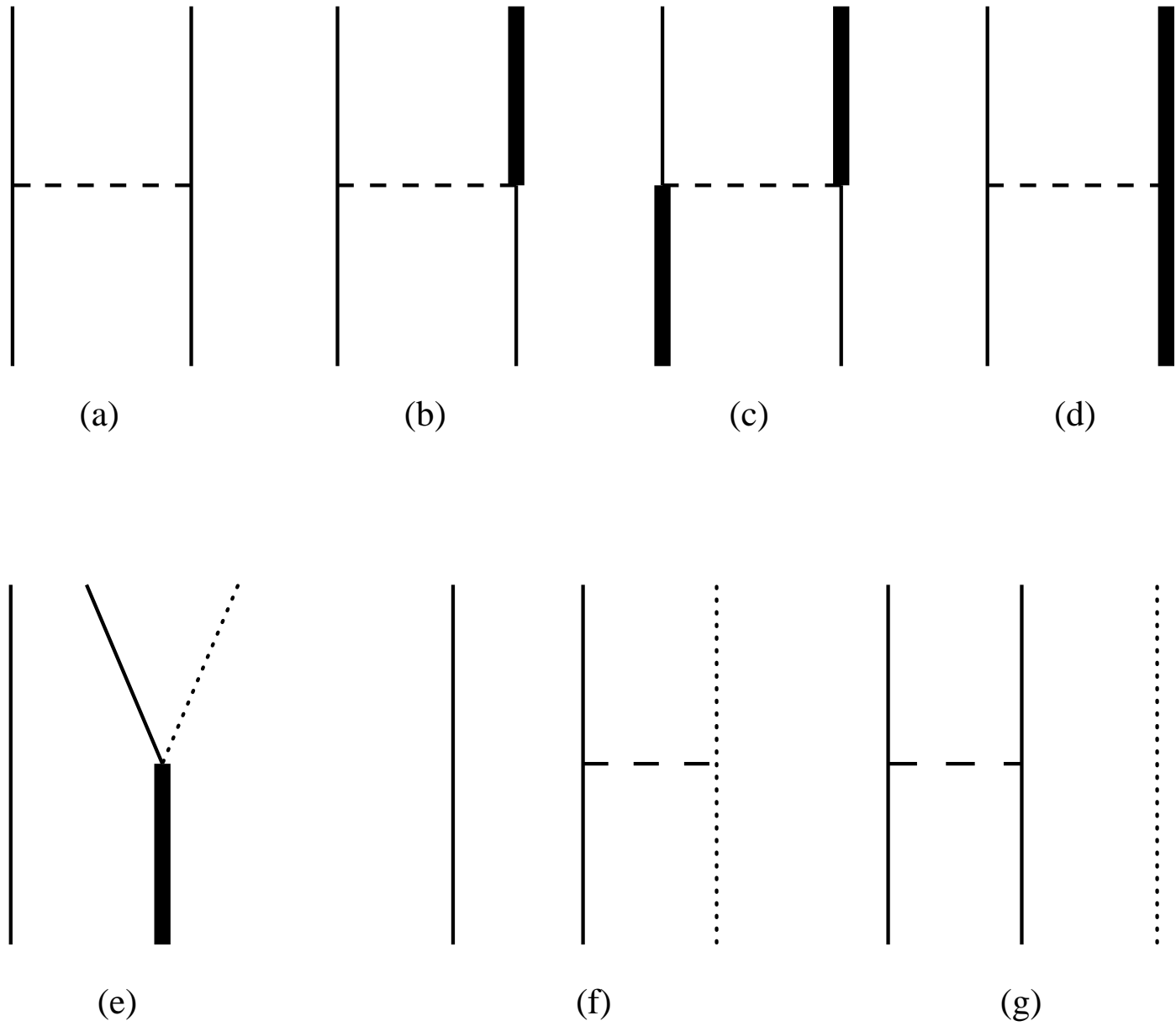
\begin{figure}
\caption[]{Graphical definition of the employed interaction hamiltonian $H_1$
for a two-baryon system. The potentials are instantaneous, the dashed lines
represent two-particle interactions in contrast to the instantaneous one-baryon
vertex process (e). Processes
(a)-(d) denote the potentials between baryons; processes (e)-(g) the coupling
to and the interaction in the Hilbert sector with a pion. The hermitian 
adjoint pieces corresponding to processes (b) and (e) are not shown.
The defined hamiltonian is an extension of a purely nucleonic one in isospin 
triplet partial waves; in isospin singlet partial waves only the purely 
nucleonic process (a) survives.\label{fig2}}
\end{figure}

The paper compares two parameterizations of the $\pi N$ $P_{33}$ resonance.
Both parameterizations are valid practical realizations of the $P_{33}$
$\pi N$ interaction in a hamiltonian with
nucleon, $\Delta$-isobar and pion degrees of freedom; the hamiltonian is
diagrammatically defined in Fig.~\ref{fig2}. Both parameterizations are valid
ones, since they account for the $\pi N$ $P_{33}$ phase shifts in comparable
quality as Fig.~\ref{fig4} and Table \ref{tab1} prove. The parameterization (P)
puts the $\pi N$ background potential to zero, the parameterization (KB) 
employs a non-vanishing one. Though the background potential is weak,
$\Delta$-isobar parameters are quite different, and, as a
consequence, the self-energy corrections of the $\Delta$-isobar in the nuclear
medium are of entirely different size, being much larger over a wide range of
energies. The latter fact is demonstrated in
Fig.~\ref{fig5}.

The two parameterizations of the $\pi N$ $P_{33}$ resonance are compared in
their effects on observables of the two-nucleon system above pion threshold and
on properties of the three-nucleon bound state. Sensitivity with respect to the 
parameterizations is clearly seen, but it is less spectacular than expected 
from the dramatic differences in the self-energy corrections of the 
$\Delta$-isobar. The inclusion of the background potential often increases the
disagreement between experimental data and theoretical prediction, especially
for elastic pion-deuteron scattering and for the pion production reaction
$pp\to\pi^+d$. Compared with the case of vanishing background, the mechanism
for pion production and pion absorption is obviously weakened in effective
strength, a net result arising from two opposing trends:
\begin{itemize}
\item The increased effective mass $M_\Delta(z,k_\Delta)$ of the
$\Delta$-isobar inhibits the $\Delta$-isobar propagation as energetically less
favorable.
\item The increased cutoff mass $\Lambda$ favors the coupling of pion-nucleon
states to the $\Delta$-isobar over a wider range of momenta.
\end{itemize}
In two-nucleon scattering above pion threshold the first trend seems to
dominate. In the three-nucleon bound state simulteneous working of both trends
is observed: The $\Delta$-isobar probability $P_\Delta$ in the wave function is
decreased, the pion probability $P_\pi$ is increased.

The two parameterizations of the $\pi N$ $P_{33}$ resonance are considered
valid ones for pion-nucleon scattering. Possibly, they could be 
differentiated and
one or the other could be ruled out, when applied to the description of
electromagnetic pion production and compton scattering on the nucleon.
Furthermore, the discouraging poor description of the two-nucleon system above
pion-threshold calls for an overall fit of the employed hamiltonian, i.e., also
of the two-baryon potentials, to the data of two-nucleon scattering and of the
unitarily coupled processes with one pion. We consider this an important,
though scaringly complicated task.
\acknowledgments
The results of the paper are based on the Diploma Thesis of G. K. concluded at
the University of Hannover in 1993.
G. K. thanks A. Valcarce who aquainted him with the techniques for carrying
out the calculations of Sect.~\ref{sec2} and \ref{sec3}, and R. W. Schulze who
was always open for conceptual questions, and who together with K. Chmielewski
provided the
code for changing the angular momentum coupling of three-baryon wave functions
between different coupling alternatives.
This work was funded by the Deutsche
Forschungsgemeinschaft (DFG) under Contract No. Sa 247/7-2 and Sa 247/7-3
(A. St.), by the Deutscher Akademischer Austauschdienst (DAAD) under Contract 
No. 322-inida-dr (T. P.), by the DOE under Grant No. DE-FG05-88ER40435 
(A. St.),
by JNICT under Contract No. PBIC/C/CEN/1094/92, and by the
Studienstiftung des deutschen Volkes (G. K.).
The numerical calculations were performed at the Regionales Rechenzentrum 
f\"ur Niedersachsen (Hannover), at the Continuous Electron Beam Accelerator
Facility (Newport News), at the National Energy Research Supercomputer
Center (Livermore), and at the National Superconducting Cyclotron Laboratory
(East Lansing).

\begin{table}
\caption[]{Parameters of the $\pi N$ hamiltonian
$(P_\Delta+Q)h(P_\Delta+Q)$ resulting from the fits of $P_{33}$ $\pi N$ phase
shifts. The first two columns 
(P) and (Pa) refer to the hamiltonian 
without $\pi N$ background potential, the version (P) was employed
in Refs.~\protect\cite{1,3,4,2}; the
columns three and four (KB) and (K) refer
to the hamiltonian with $\pi N$ background,
developed in this paper. Column one, labelled (P), repeats the parameters of
Ref.~\protect\cite{3}, obtained under the assumption of a resonance position
$m_Rc^2$ at 1236 MeV. In column two, labelled (Pa), the hamiltonian is adapted
to the improved experimental data of Ref.~\protect\cite{5} with
a resonance position $m_Rc^2$ of 1232 MeV. Column three, labelled (KB), lists
the parameters for the hamiltonian of this paper. The last row indicates the 
quality of the achieved fits by $\chi^2/N$, $N=28$, with respect to the data of 
Ref.~\protect\cite{5}.
Since error bars are not given for the ``experimental'' phase shifts of
Ref.~\protect\cite{5}, ``experimental'' uncertainties of $1^{\circ}$ are 
assumed
for all of them when calculating $\chi^2/N$. The set of parameters in column
four, labelled (K), is only used when in a calculation with the full
hamiltonian (KB) the pure resonance contribution to an observable is to be 
isolated.
It reproduces the correct resonance position, though. The parameter set of 
column
four does not constitute a valid parameterization of the hamiltonian by itself,
the resulting $\chi^2/N$ is very poor, though not
outrageously wrong. The dashed line of Fig.~\protect\ref{fig4} reflects that
fact.}
\begin{tabular}{c|cc|c|c}
&Ref.~\protect\cite{3}&adapted&\\
&(P)&(Pa)&(KB)&(K)\\
\tableline
$m_Rc^2$ [MeV]&1236.0&1232.0&1232.0&1232.0\\
$m_\Delta^0c^2$ [MeV]&1315.0&1311.0&1801.0&1801.0\\
$\Lambda$ [MeV/c]&287.9&287.9&859.36&859.36\\
$\frac{f^2}{4\pi}\frac1{(\hbar c)^3}$&0.306&0.306&0.306&0.306\\
$\lambda_1$ [1/MeV]&0&0&-0.0522&0\\
$\lambda_2$ [1/MeV]&0&0&0.273&0\\
$\hbar\beta_1$ [MeV/c]&-&-&369.1&-\\
$\hbar\beta_2$ [MeV/c]&-&-&597.76&-\\
\tableline
$\chi^2/N$&10.0&1.7&0.8&76.4
\label{tab1}
\end{tabular}
\end{table}

\begin{table}
\caption[]{Results for some trinucleon bound state properties. 
Results, based on the two parameterizations (P) and (KB) of the $P_{33}$
$\pi N$ interaction, are compared; the results for (P) are identical with
those of Ref.~\protect\cite{2} labelled $H(1)$ there.
The table lists the triton binding energies $E_T$, binding energy corrections
arising von non-nucleonic degrees of freedom in the definition of
Ref.~\protect\cite{11}, $\Delta E_2$ being the binding energy correction of
two-baryon nature; and
$\Delta E_3$ being the corresponding correction of three-baryon nature.
The table also lists the wave function probabilities, i.e., $P_{{\cal L}}$ for
nucleonic components of total orbital angular momentum ${\cal L}=S,P,D$ and
of particular orbital permutation symmetry, the probability $P_\Delta$ for
components with a $\Delta$-isobar, and the probability $P_\pi$ for components
with a pion. The binding energies in the first two columns result from exact
Faddeev calculations, they are correct within 10 keV only, 
but the last digits in
rows $E_T$, $\Delta E_2$ and $\Delta E_3$ are believed to represent 
relative changes between the parameterizations correctly. The binding energy
correction of first order in $Qt_{BG}(z)Q$ in the third column is derived in 
perturbation theory according to Eq.~(\ref{4.2c}).}
\begin{tabular}{l|c|cc}
&(P)&\multicolumn{2}{c}{(KB)}\\
$m_\Delta^0$ [MeV/$c^2$]    &1315.0 &\multicolumn{2}{c}{1801.0} \\
&&zeroth order&first order\\
&&in $Qt_{BG}(z)Q$&in $Qt_{BG}(z)Q$\\
\tableline
$E_T$[MeV]       &-7.849 &-7.731 &-7.730 \\
$\Delta E_2$[MeV]& 0.456 & 0.376 & 0.372 \\
$\Delta E_3$[MeV]&-0.924 &-0.726 &-0.721 \\
$P_S$[\%]        & 88.23 & 88.70 & \\
$P_{S'}$[\%]     & 1.24  & 1.27  &  \\
$P_P$[\%]        & 0.08  & 0.08  &  \\
$P_D$[\%]        & 8.68  & 8.59  &  \\
$P_\Delta$[\%]   & 1.71  & 1.16  &  \\
$P_\pi$[\%]      & 0.06  & 0.195 &  \label{tab2}
\end{tabular}
\end{table}

\begin{figure}
\caption[]{Characteristic contributions to the $P_{33}$ $\pi N$ transition 
matrix. Process (a) is a purely resonant process, it does not contain any 
background contribution, while process (b) is a pure background interaction.
Process (b)
is an example of how the background potential contributes to the $P_{33}$ 
$\pi N$ scattering; it represents a series of processes in which the potential
$Qh_1Q$ is to be replaced by the ladder sum of the transition matrizes
$Qt_{BG}(z)Q$.
The processes (a) and (c) are sample processes contained in the definition 
(\protect\ref{2.5}) of $\Delta$-isobar self-energy corrections.
\label{fig3}}
\end{figure}

\begin{figure}
\caption[]{$\pi N$ phase shifts in the $P_{33}$ partial wave. 
The results for different parameterizations
of the $P_{33}$ resonance are compared. The diamonds are the experimental data 
points from \protect\cite{5} used for the fit of this paper;
the diagonal crosses represent newer experimental data according to
Ref.~\protect\cite{8}, which, however, are not taken into account for the
present work. The parameterization (KB) of this paper for the 
$P_{33}$ $\pi N$ interaction
with background potential is shown as solid curve. The parameterization (Pa)
without background potential is an improvement of the version (P) given
in Ref.~\protect\cite{3}; it is shown as dotted curve.
The dashed
line shows the resonance contribution of the parameterization (KB) alone; the
corresponding parameters are collectively labelled (K) in 
Table~\protect\ref{tab1}; the parameters (K) {\it do not} constitude a valid
parameterization of $P_{33}$ $\pi N$ scattering by themselves.
\label{fig4}}
\end{figure}

\begin{figure}
\caption[]{The effective mass $M_\Delta(z,k_\Delta)$ and width 
$\Gamma_\Delta(z,k_\Delta)$ of the $\Delta$-isobar, as defined in
Eqs.~(\protect\ref{2.6a}) and (\protect\ref{2.6b}), respectively. 
Their dependence on the available energy $z$ is shown
for vanishing $\Delta$-momentum, i.e., for $\hbar k_\Delta=0$. Results for the 
parameterization (KB)
of this paper with a non-vanishing $\pi N$ background potential 
and for the adapted parameterization (Pa) with vanishing 
$\pi N$ background potential are compared by solid and dotted lines,
respectively. The two compared parameterizations have the 
bare $\Delta$-masses 1801 MeV$/c^2$ and 1311 MeV$/c^2$.
\label{fig5}}
\end{figure}

\begin{figure}
\caption[]{Effective $N\Delta$ interactions. The processes (a) and (b) are of
one-baryon nature, the processes (c)-(f) of two-baryon nature. Only the
processes (a) and (c) survive in case the background potential is
assumed to vanish. The processes (b), (d) and (e) are first order in the
background potential $Qh_1Q$, process (f) of second order. Each of the
processes (b), (d)-(f) represents a series of processes in which the potential
$Qh_1Q$
is to be replaced by the ladder sum of the transition matrix $Qt_{BG}(z)Q$.
\label{fig6}}
\end{figure}

\begin{figure}
\caption[]{${}^1D_2$ phase shifts and inelasticities of elastic two-nucleon
scattering as a function of the nucleon lab energy. Results
for the parameterization (P) without background potential, and for
the parameterization (KB) with background potential are shown as dotted and
solid curves. The effect of the background potential is mostly indirect: It
changes the bare $\Delta$-mass $m_\Delta^0$ and the parameters of the $\pi
N\Delta$ vertex. The direct influence of the background potential is omitted in
the results of the dashed curve --- it is based on the parameterization (KB),
but omitting all background contributions.
The experimental data are taken from the energy-independent phase shift 
analysis
of Ref.~\protect\cite{6}.
\label{fig7}}
\end{figure}

\begin{figure}
\caption[]
{Differential cross section for pion production in $pp\to\pi^+d$ at two
proton lab energies as function of the pion scattering angle in the $\pi d$
c.m. system.
Results for the parameterization (KB) with background potential and for the
parameterization (P) without background potential are compared as solid and
dotted curves. The data are taken 
from the compilation of 
Ref.~\protect\cite{7}.\label{fig8}}
\end{figure}

\begin{figure}
\caption[]{Differential cross sections for elastic pion deuteron scattering
at two pion lab energies as a function of the pion scattering angle in the
$\pi N$ c.m. system. Results for the parameterization (KB) with background 
potential and for the
parameterization (P) without background potential are compared as solid and
dotted curves. 
The data are taken from Ref.~\protect\cite{9,10}.\label{fig9}}
\end{figure}

\begin{figure}
\caption[]{Examples for the effective three-baryon interaction in the Hilbert
sector ${{\cal H}}_\Delta$ arising from the $\pi N$ background potential. The
processes (a), (b) and (c) are of first, second and third order in the
background potential $Qh_1Q$. Each of the processes represent a series of
processes in which the potential is to be replaced by the ladder sum of the
transition matrix $Qt_{BG}(z)Q$. Even in this extended form, the shown five
processes represent only the lowest order ones of the Faddeev-Yakubovsky series
\protect\cite{jakob} for
four particles interacting through potentials of very restrictive character.
$NN$ interactions within the pionic Hilbert sector ${{\cal H}}_\pi$ are not
considered here, even though they would, through processes like
process (d) and (e), also give rise to an effective three-baryon interaction 
in the Hilbert sector
${{\cal H}}_\Delta$; the appendix of Ref.~\cite{2} describes the technical
treatment of the disconnected process (d); process (e) is fully connected.
\label{fig10}}
\end{figure}
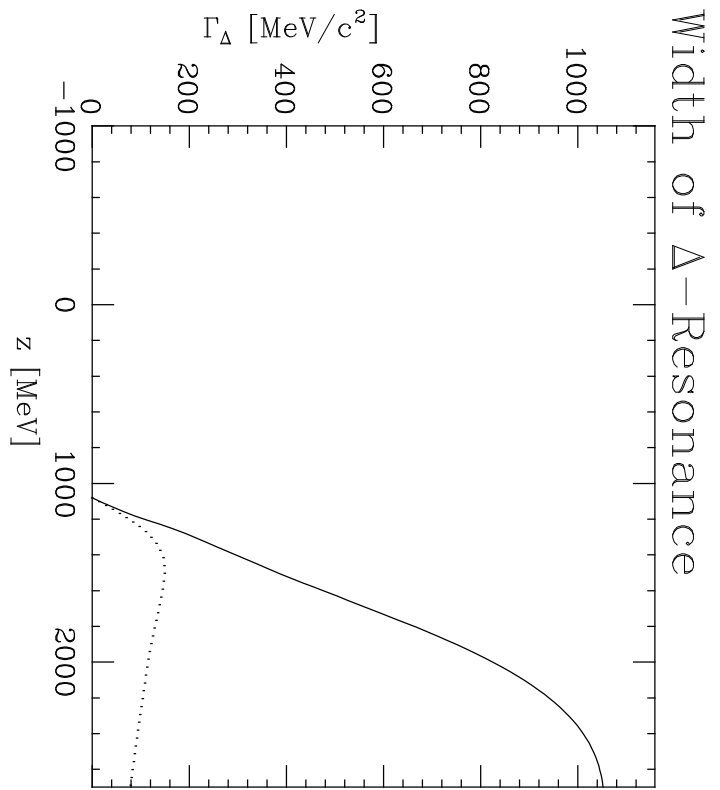
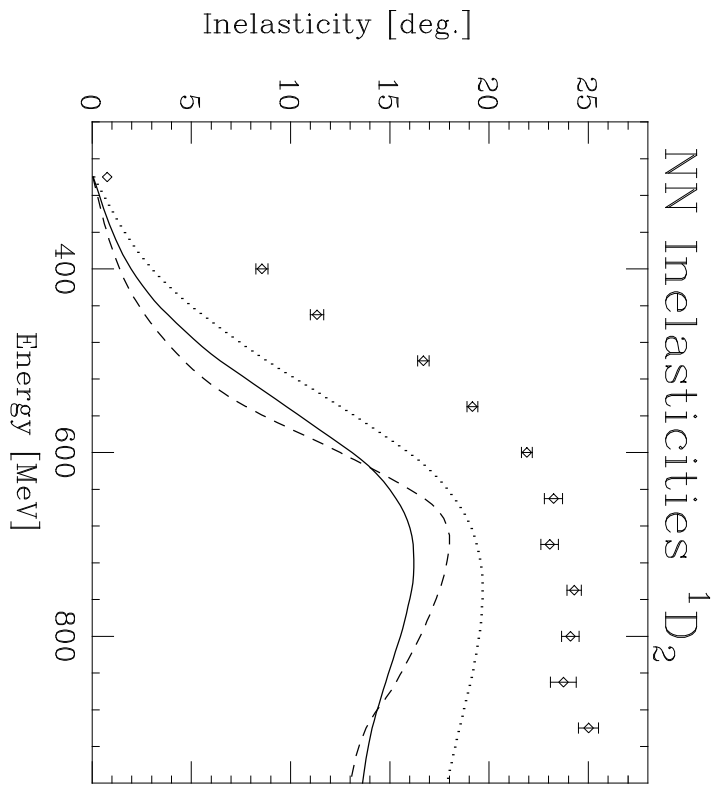
\end{document}